\documentclass[prb,aps,twocolumn,amsmath,amssymb,floatfix,
superscriptaddress]{revtex4-1}
\DeclareUnicodeCharacter{2009}{\,}
\usepackage[dvips]{graphics}
\usepackage{graphicx}
\usepackage{color}
\usepackage{blindtext}
\definecolor{dred}{rgb}{0.75,0,0}
\usepackage{graphicx}
\usepackage{dcolumn}
\usepackage{bm}
\usepackage{xcolor}
\usepackage{braket}
\usepackage{physics}
\usepackage{appendix}
\usepackage{braket}
\usepackage{hyperref}
\hypersetup{
 colorlinks=true,
citecolor=magenta,filecolor=blue,
 linkcolor=magenta,
 }

\usepackage{graphicx}
\usepackage{dcolumn}
\usepackage{bm}
\usepackage{xcolor}

\definecolor{codegreen}{rgb}{0,0.6,0}
\definecolor{codegray}{rgb}{0.5,0.5,0.5}
\definecolor{codepurple}{rgb}{0.58,0,0.82}
\definecolor{backcolour}{rgb}{0.95,0.95,0.92}

\begin{document}

\preprint{APS/123-QED}

\title{\textcolor{blue}{Flat bands and topological phase transition in entangled Su-Schrieffer-Heeger chains}} 

\author{Sauvik Chatterjee}
\affiliation{Department of Physics, Presidency University, 86/1 College Street, Kolkata, West Bengal - 700 073, India}
\affiliation{chatterjeesauvik8@gmail.com}
\author{Sougata Biswas}
\affiliation{Department of Physics, Presidency University, 86/1 College Street, Kolkata, West Bengal - 700 073, India}
\affiliation{sougata.rs@presiuniv.ac.in}
\author{Arunava Chakrabarti}
\affiliation{Department of Physics, Presidency University, 86/1 College Street, Kolkata, West Bengal - 700 073, India}
\affiliation{arunava.physics@presiuniv.ac.in}
\date{\today}

\begin{abstract}
Flat, non-dispersive bands and topological phase transition in multiple Su-Schrieffer-Heeger (SSH) chains, cross-linked via periodically arranged nodal points are explored within a tight binding framework. We give an analytic prescription, based on a real space decimation scheme, that extracts the energy eigenvalues corresponding to the flat bands along with their degeneracy. The topological phase transition is confirmed through the existence of quantized Zak phase for all the Bloch bands, and the edge states that are protected by chiral symmetry, consistent with the bulk-boundary correspondence. In addition to the edge states, the entangled systems are shown to give rise to clusters of localized eigenstates in the bulk of the system, in contrast to a purely one-dimensional SSH system.
\end{abstract}

\maketitle

\section{Introduction}
\label{intro}
The paradigmatic Su-Schrieffer-Heeger (SSH) model~\cite{su,heeger}, described within a tight binding framework, has been instrumental in understanding the conjugated polymers~\cite{yulu,baeriswyl}. The SSH Hamiltonian is described by a staggered distribution of two `hopping integrals' ($v$ and $w$, say) and mimics the alternating bond pattern of a polyacetylene. The primary and the most important cause behind the huge attention paid to this model in recent years has been the cause that, the bipartite sublattice structure of the model is emblematic of a one-dimensional chiral symmetric analogue of the topological insulators~\cite{thouless} that have been at the center stage of a broad area of recent research in condensed matter physics~\cite{kane,bernevig,fu}. The most important support to the topological properties of an SSH chain is given by a topological invariant in the form of a quantized value of the so-called Zak phase~\cite{zak}, and a symmetry protected, robust edge state~\cite{asboth} that appears as the gaps in the Brillouin zone boundaries open up on tuning the values of the staggered hopping integrals. 

Such a simple toy model, yielding a wealth of new physics, has triggered intense research activity on different extensions or variants of the basic SSH structure. The reason is simple - to check whether the fundamental physics embedded in the SSH model remains intact on its non-trivial extensions. For example, some of the interesting results include studies of the topological properties of two coupled SSH chains~\cite{li}, SSH chains with long-range interaction~\cite{chun-fang}, extended SSH models with non-local couplings~\cite{miroshnichenko}, observation of topologically protected edge states in a trimer-lattice~\cite{martinez} or a four-bond SSH model~\cite{bid} to name a few. A one-dimensional array of diamond-shaped loops threaded by a staggered SSH-like distribution of trapped magnetic flux has been examined in search of edge states with topological protection~\cite{amrita1}. The topology of multi-strand Creutz ladder networks has also been explored to work out the energy bands and the topological invariants~\cite{amrita2}.  The field is enriched by studies such as examining the influence of an absence of a non-centered inversion symmetry~\cite{ricardo1}, exploring the topological properties of two bosons in a flat-band system~\cite{ricardo2} or extending the ideas of the topological phase transitions to a square-root model on a photonic lattice~\cite{alex} or to the $2^n$-root topological insulators~\cite{marques}. These have indeed widened the scope of research in this field.

\begin{figure}[ht]
(a)\includegraphics[width=.9\columnwidth]{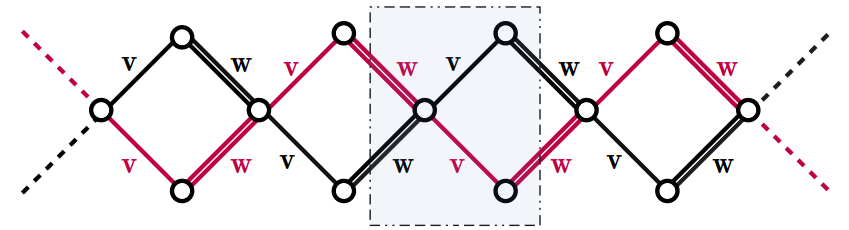}
(b)\includegraphics[width=.9\columnwidth]{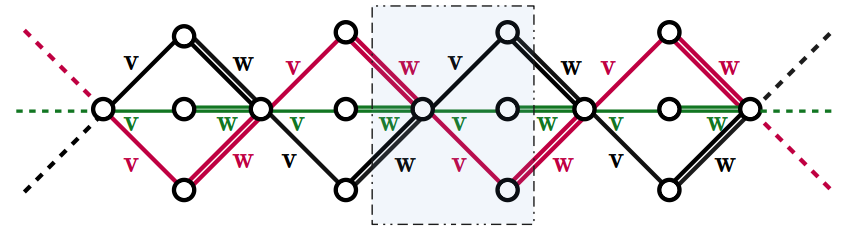}
(c)\includegraphics[width=.9\columnwidth]{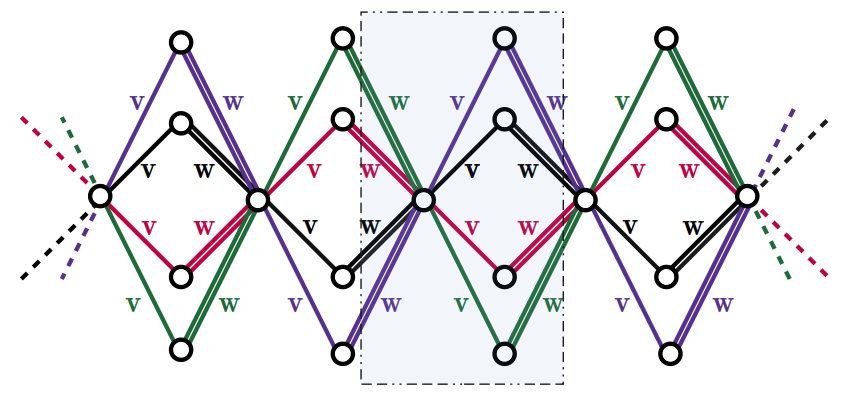}

\caption{(Color online) Examples of cross-linked SSH chains. From (a) to (c) the number of chains stitched together at the nodal points are two, three and four respectively. The hopping integrals along every chain alternate with values $v$ and $w$ respectively, as shown. }
\label{fig1}
\end{figure}

In a very recent work~\cite{sivan} an interesting model has been proposed and analysed, in which several SSH chains are stitched at one point, the linking point serving the role of an `impurity'. The existence of cross-linking sites in polymer chains is an inspiration behind such an investigation. The edge states, their protection, and the localization properties have been discussed, and some interesting aspects have been reported.

Motivated by this observation we undertake an in-depth analysis of several SSH chains, each with the identical distribution of the staggered hopping amplitudes, entangled in every unit cell (Fig.~\ref{fig1}). The construction mimics, though in an elementary way, several polymer chains winding periodically along the $x$-axis. The number of SSH chains stitched together at the nodes can be anything between two and infinity (in principle), rendering the coordination number of the `cross-linking' points values between two and infinity. This is equivalent to a {\it local} geometric `disorder', though in each case the disordered environment is repeated periodically along the major axis of the model polymer. Does it alter, in any way, the basic topological properties exhibited by a single SSH chain? This is the question we ask ourselves and try to find an answer to in this work. 

It is pertinent to address here the issue related to the possible experimental realization of such one or quasi-one-dimensional models. Several path breaking experiments in recent times have indeed been inspiring in this regard. To cite a few examples, the experimental realization of one-dimensional topological edge states observed on Indium atomic wires self-assembled on a silicon surface has been reported~\cite{cheon}. A direct fabrication of nanowires with less than a nanometer width and Y junctions connecting designated points in a transition-metal dichalcogenide monolayer is possible these days~\cite{lin}. Even, an experimental development of an SSH system with multiatomic base in transition metal monochalcogenide nanowires has confirmed topological features~\cite{jin-liu}. Atomically controlled trimer and coupled dimer chains have demonstrated the topological domain wall excitation modes using a scanning tunneling micriscope~\cite{nurul}.

The results we get in the present work are interesting. Firstly, the topological phase transition is indeed seen, in each case, to occur exactly under the same conditions as that of a single unperturbed SSH chain. The Zak phase is exactly obtained and is seen to flip its value from $0$ to $1$ under appropriate conditions signifying a change in the topological state of the system. The localized states, which are all found to be protected by chiral symmetry, are found to exist at one edge, consistent with the bulk-boundary correspondence. In addition to the edge states, we find localized states to exist even in the bulk of the sample, with different localization lengths. This effect is attributed to the existence of the nodal points, as mentioned in the text.

Secondly, the looped structures of the assembly of the SSH chains give rise to flat, non-dispersive energy bands~\cite{leykam,flach,bodyfelt,xia}. The flat bands (FB) have been experimentally engineered on a variety of photonic lattices~\cite{camilo,weiman,aravena,salinas,matheus} and their role in controlling the transport or the prospect of transferring the photonic modes between topological edge states have been discussed in the literature. The FB's we extract here are found to be $N-1$ fold degenerate for a system of $N$ SSH chains cross-linking together. We present an exact analytical way to discern such flat band (FB)-energies, and most importantly, their degeneracies. The analytical results are corroborated by the results obtained by a direct diagonalization of the Hamiltonians in each case, written in the momentum space. The match between the two methods is exact.

We now try to elucidate our findings. While doing so, we choose to talk about the occurrence and the degeneracy of the FB's first, and describe a real space decimation method to unravel the FB's in the entangled SSH loops. This is done in section II. In section III we discuss the topological phase transition of the entangled systems and in section IV we draw our conclusions.
\begin{figure}[ht]
\includegraphics[width=.9\columnwidth]{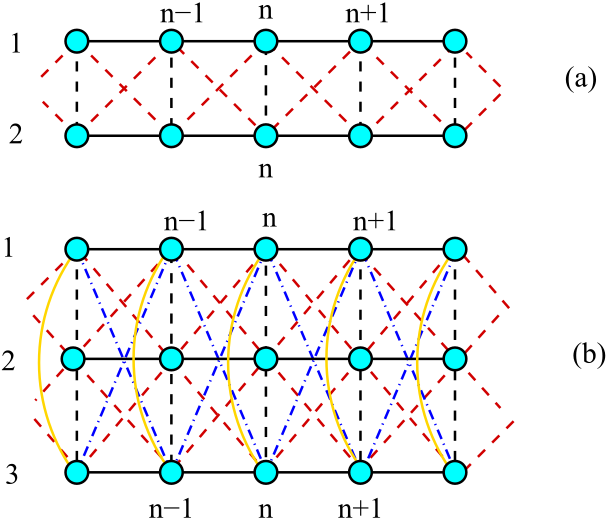}
\caption{(Color online) The entangled SSH chains corresponding to Figs.~\ref{fig1}(a) and (b), mapped onto a two-strand (a) and a three-strand (b) ladder network respectively,  by decimating out the cross-link nodes with coordination numbers $4$ and $6$ in Fig.~\ref{fig1} (a) and (b) respectively. The decimation leads to longer-range energy dependent hopping integrals, depicted by different colors. The explicit values of the renormalized on-site potential and the hopping integrals are given in the text.}
\label{ladder}
\end{figure}
\section{Flat bands in an entangled SSH polymeric system}

\subsection{The Hamiltonian}

\begin{figure*}[ht]
\centering
(a)\includegraphics[width=0.6\columnwidth]{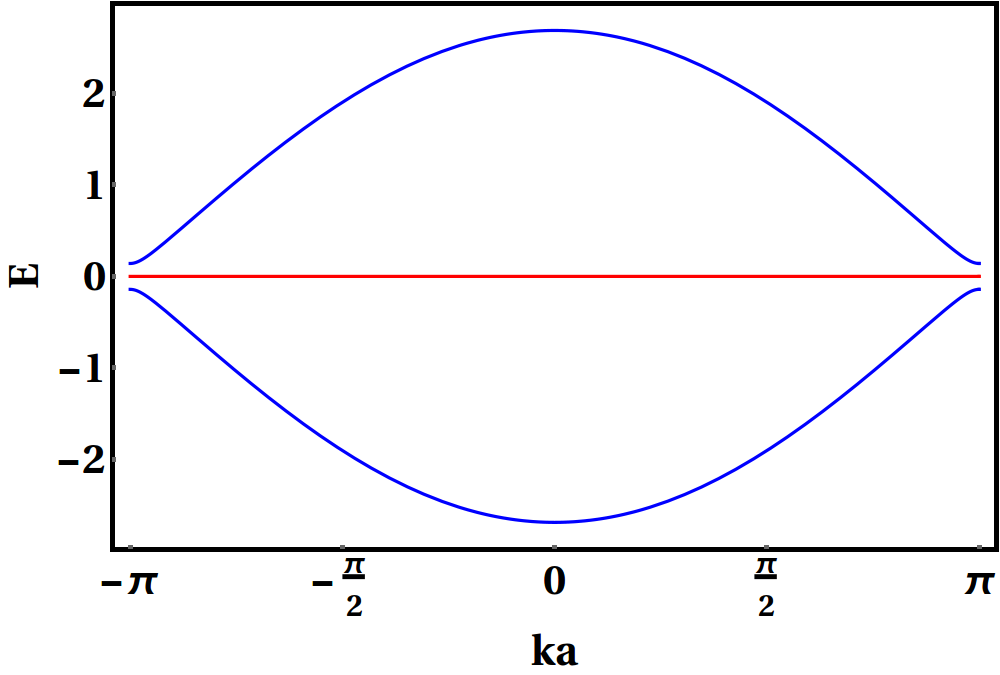}
(b)\includegraphics[width=0.6\columnwidth]{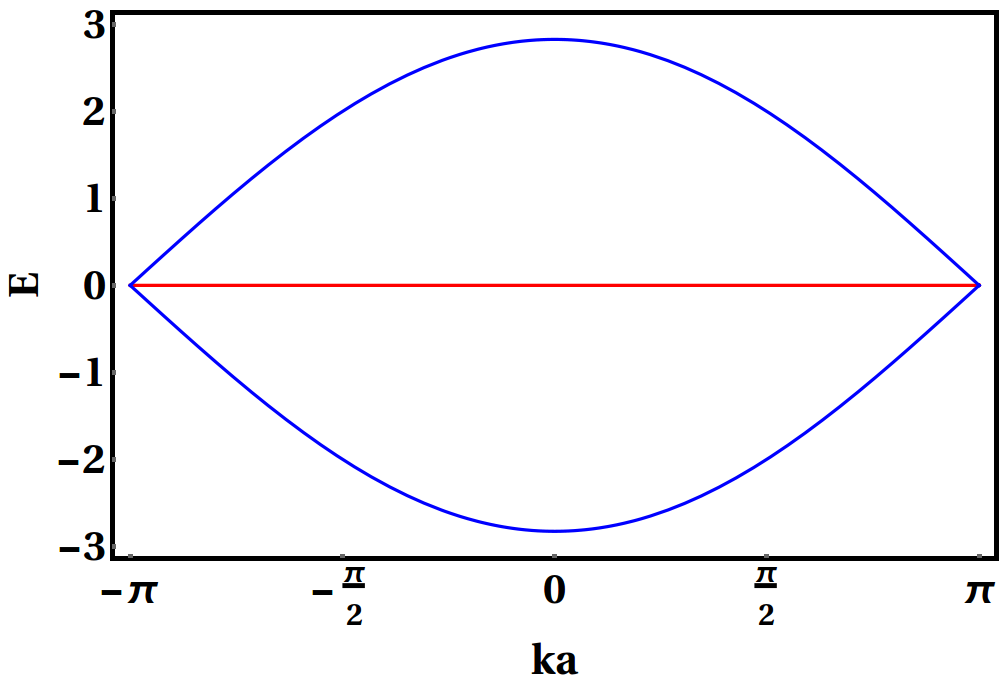}
(c)\includegraphics[width=0.6\columnwidth]{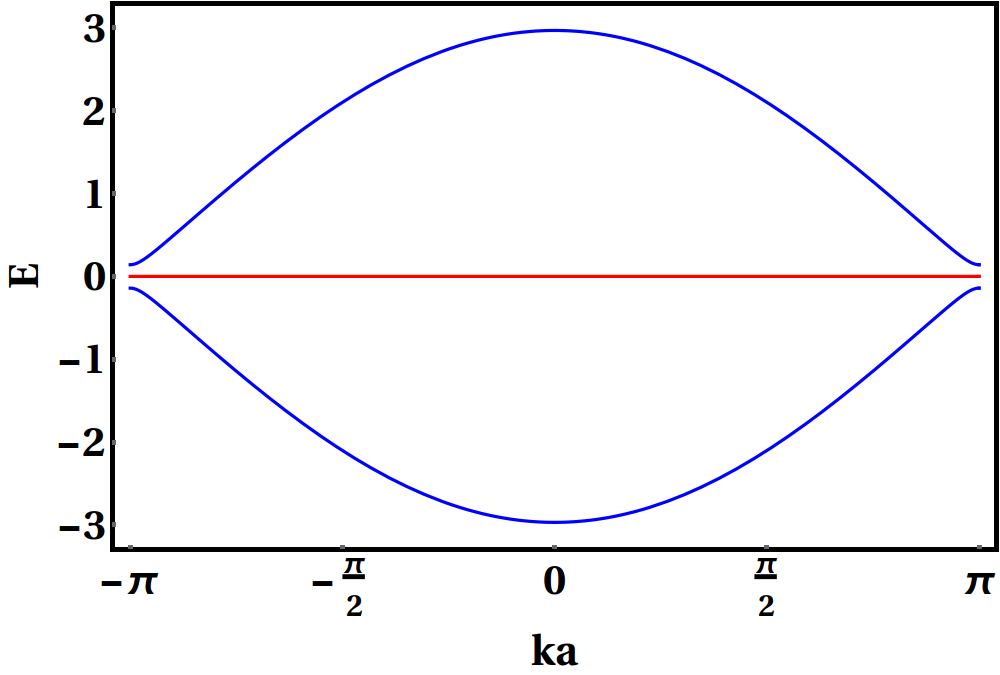}
(d)\includegraphics[width=0.6\columnwidth]{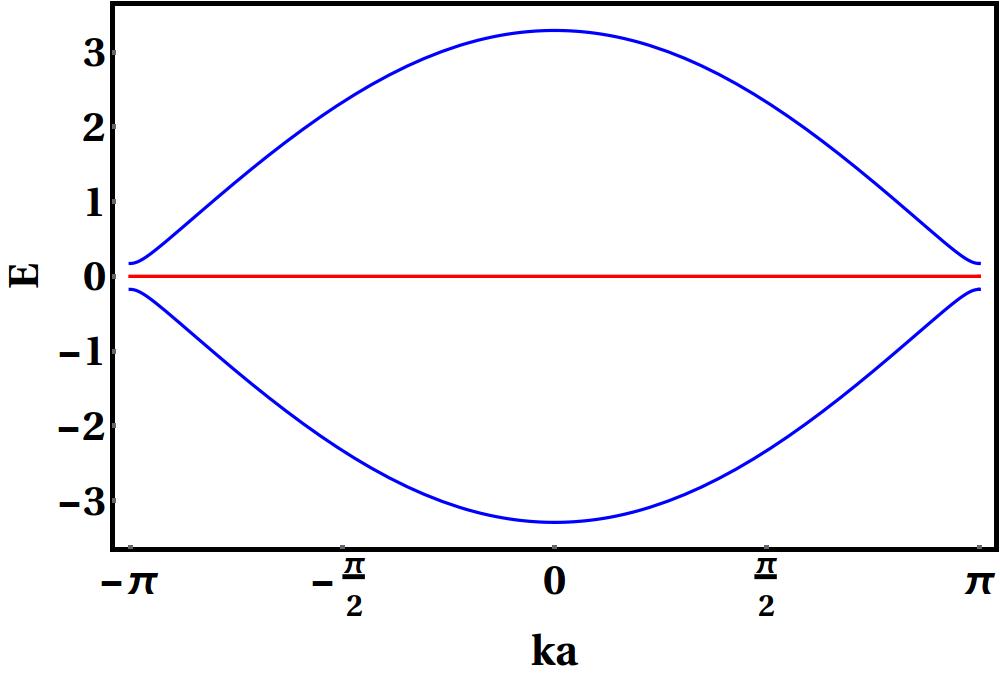}
(e)\includegraphics[width=0.6\columnwidth]{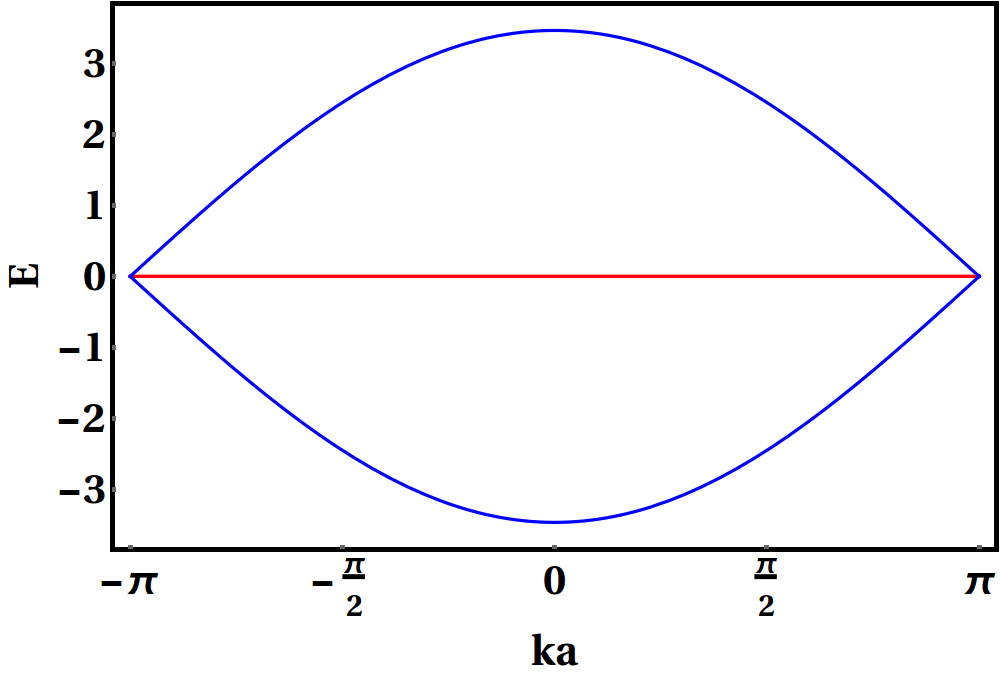}
(f)\includegraphics[width=0.6\columnwidth]{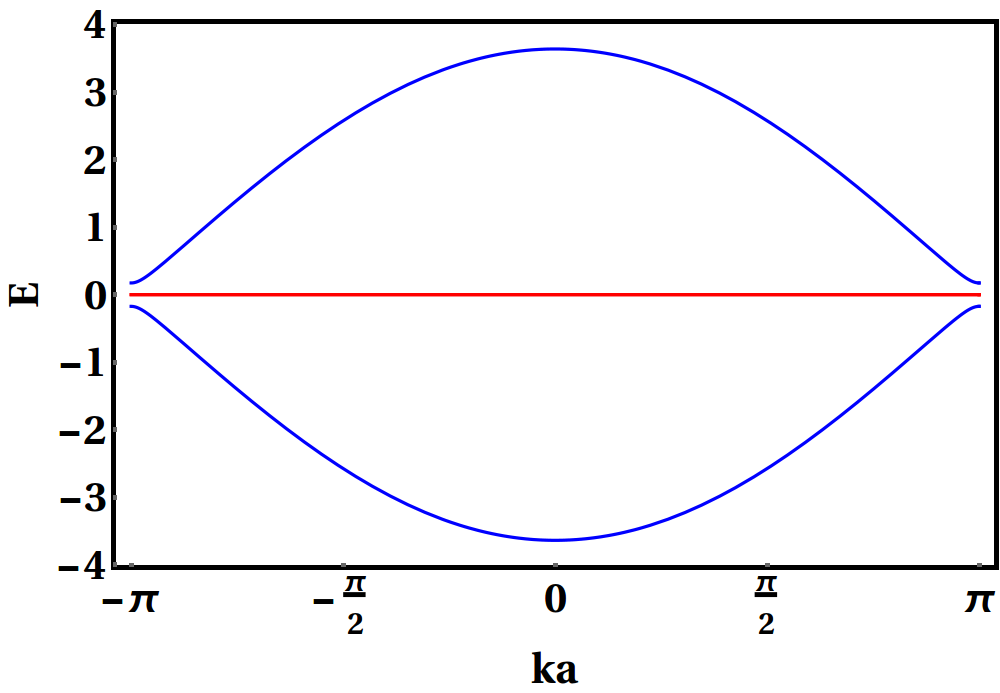}
(g)\includegraphics[width=0.6\columnwidth]{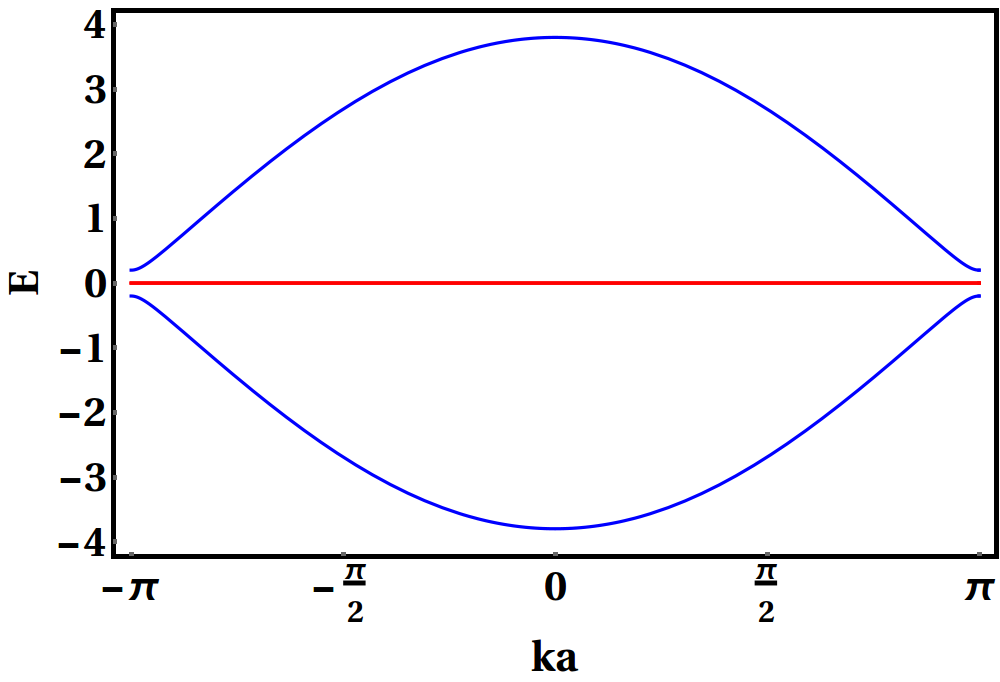}
(h)\includegraphics[width=0.6\columnwidth]{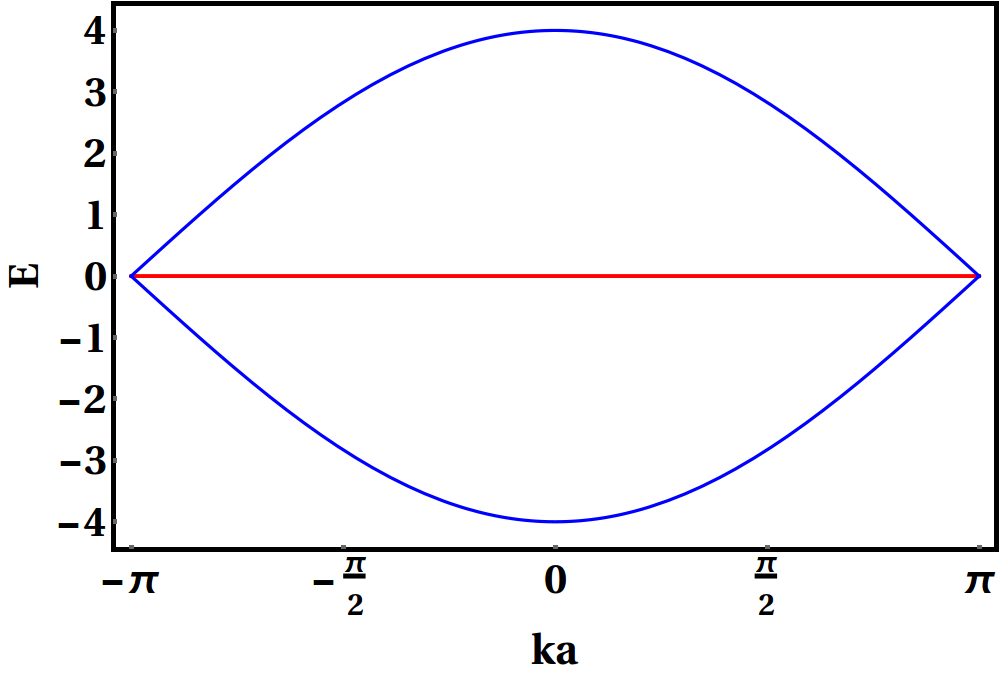}
(i)\includegraphics[width=0.6\columnwidth]{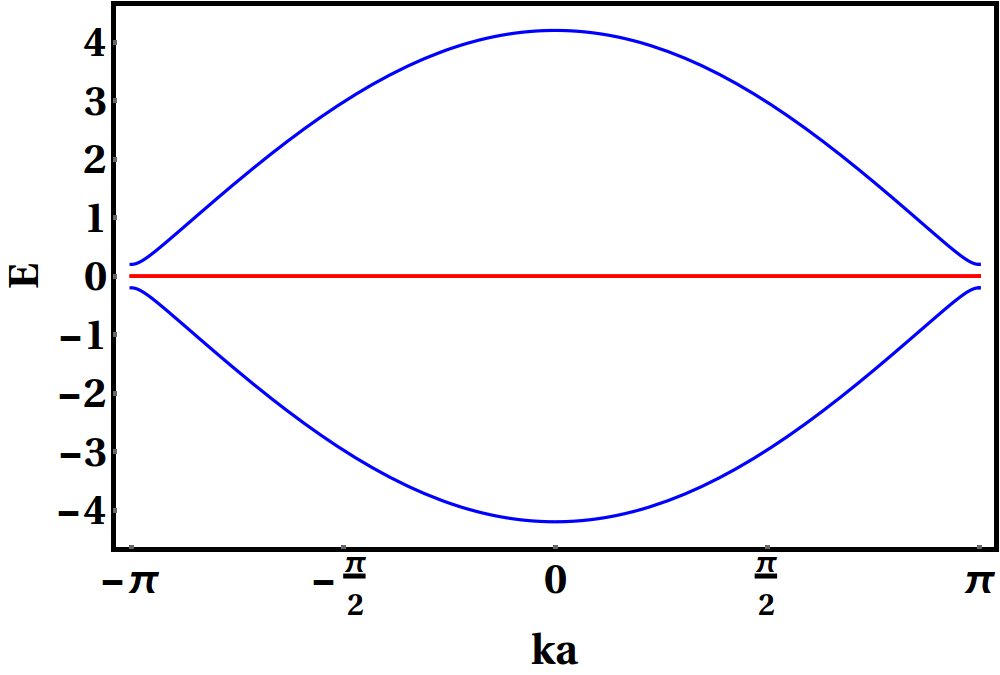}
\caption{(Color online) Energy vs wave vector (E vs ka) dispersion relations for two cross-linking SSH chains ((a),(b),(c)), three cross-linking SSH chains ((d),(e),(f)), and four cross-linking SSH chains ((g),(h),(i)) respectively, as shown in  Fig.~\ref{fig1}. The parameters are chosen as $\epsilon = 0 $,  $v=1.0$, $w=0.9$ for (a), (d) and (g), $\epsilon = 0 $, $v=1.0$, $w=1.0$ for (b), (e) and (h), and  $\epsilon = 0 $,  $v=1.0$, $w=1.1$ for (c), (f) and (i) respectively. In the two-chain case, a single flat band is present, while in the three-chain and four-chain cases, there are doubly and triply degenerated flat bands are observed respectively, and are marked by red color.}  
\label{disp1}
\end{figure*}

Our model systems are shown in Fig. \ref{fig1} (a)-(c). Identical SSH chains cross each other at the nodal points as shown. Each chain is colored separately. The coordination number of the nodal point can take any value between $0$ and $\infty$ (in principle). Needless to say that, from a nodal point an electron can hop along any other chain (with a different color) and yet can feel the same SSH environment as its parent chain. We show just three illustrative examples here. The basic form of the tight binding Hamiltonian remains as usual, that is, 
\begin{equation}
    H = \sum_i  \epsilon_i \ket{c_i} \bra{c_i} + \sum_{<ij>} 
    ( \ket{c_i} t_{ij} \bra{c_j} + h.c )
    \label{ham}
\end{equation}
Here, $\epsilon_i$ in the `on-site' potential at the $i$-th atomic site, is eventually taken to be constant throughout and set equal to zero in this work. $t_{ij}$ represents the amplitude of the nearest neighbor hopping integral, and in this work, we choose $t_{ij} =v$, and $w$, two real numbers arranged periodically along any of the  SSH chains that cross-link. 

To unravel the FB's and the topological states, we use the discrete version of the Schr\"{o}dinger equation, viz, the `difference equations' that form a set of linear algebraic equations,

\begin{equation}
    (E-\epsilon_j)\psi_j = \sum_{k} t_{jk}\psi_k
    \label{diffeqn}
\end{equation}
where, $\psi_j$ is the amplitude of the wavefunction on the $j$-th site, and the summation runs over the nearest neighbors of $j$. Naturally, at the linking sites we shall have more number of terms on the right hand side of Eq.~\eqref{diffeqn} compared to that for a site in an arm of the constituent SSH segment. 

\subsection{Discerning the flat bands}
We explain our scheme for the first two of the geometries in Fig.~\ref{fig1}. The third geometry in Fig.~\ref{fig1}(c) is also dealt with in a similar fashion. However, we skip the details here as the four-strand effective ladder network, in this case, looks cumbersome. 

Using the set of Eq.~\eqref{diffeqn}, we decimate out the amplitudes $\psi_j$ corresponding to the nodal points with coordination number four in Fig.~\ref{fig1}(a) and (b). The sites surviving the decimation are the cyan-colored ones. The decimation maps the figures into a two-strand and a three-strand ladder respectively, where each surviving site has an energy-dependent potential, and energy dependent hopping integrals, ranging beyond the nearest neighbor. The range of hopping increases to the second neighbor for Fig.~\ref{ladder}(a) and to the second and third neighbors for Fig.~\ref{ladder}(b). The different colors represent the different ranges of interaction on the ladder network.

The difference equations for the two cases cited above are written conveniently in matrix form. The equations are,
\vskip .2in

\begin{eqnarray}
    [E.\mathbb{I}_{2\cross2} - \Tilde{\epsilon_{2}}] \Psi_n & = & \Tilde{t_{2}}(\Psi_{n+1} + \Psi_{n-1})
  \label{eq1}
\end{eqnarray}
for the two-strand case, and, 
\begin{eqnarray}
        [E.\mathbb{I}_{3\cross3} - \Tilde{\epsilon_{3}}] \Psi_{n} & = & \Tilde{t_{3}}(\Psi_{n+1} + \Psi_{n-1}) 
    \label{eq2}
\end{eqnarray}
for the three-strand case respectively. Here, we have defined 
\begin{equation}
\Psi_{n} = \left[ \begin{array}{cccccccccccccccc}
\psi_{n,1}\\
 \psi_{n,2}
\end{array}
\right ] 
\end{equation}
and, 
\begin{equation}
\Psi_{n} = \left[ \begin{array}{cccccccccccccccc}
\psi_{n,1}\\
 \psi_{n,2}\\
 \psi_{n,3}
\end{array}
\right ] 
\end{equation}
In the expressions for the wavefunctions $\psi_{n,j}$ signifies the amplitude at the $n$-th site of the $j$-th strand of the ladder, and $\mathbb{I}_{2 \times 2}$ and $\mathbb{I}_{3 \times 3}$ are the $2 \times 2$ and $3 \times 3$ unit matrices respectively.
The decimation yields {\it renormalized} values of the on-site potentials, written here as a $2 \times 2$ matrix $\tilde{\epsilon_2}$ and a $3 \times 3$ matrix $\tilde{\epsilon_3}$ for the two-strand and the three-strand cases respectively. The renormalized hopping integrals for the ladder networks are designated by the $2 \times 2$ matrix $\tilde{t}_2$ and the $3 \times 3$ matrix $\tilde{t}_3$ respectively. The matrices now include the nearest neighbor and the longer-range hopping terms. $\tilde{\epsilon}_2$, $\tilde{\epsilon}_3$, $\tilde{t}_2$ and $\tilde{t}_3$ are written in terms of the original parameters as, 
\begin{equation}
\Tilde{\epsilon_{2}} = \left[ \begin{array}{cccccccccccccccc}
\epsilon + \Gamma & \Gamma\\
 \Gamma & \epsilon + \Gamma \nonumber\\
\end{array}
\right ] 
\end{equation}
\begin{equation}
\Tilde{t_{2}} = \tau \mathbf{\Lambda}_{2\times2} \nonumber \\
\end{equation}
and, 
\begin{equation}
\Tilde{\epsilon_{3}} = \left[ \begin{array}{cccccccccccccccc}
\epsilon + \Gamma & \Gamma & \Gamma\\
 \Gamma & \epsilon + \Gamma & \Gamma\\
 \Gamma & \Gamma & \epsilon+\Gamma\nonumber\\
\end{array}
\right ] 
\end{equation}
\begin{equation}
\Tilde{t_{3}} = \tau \mathbf{\Lambda}_{3\times 3} \nonumber \\
\end{equation}
with $\mathbf{\Lambda}_{N \times N}^{ij}= 1 $ for all $(i,j)$. 
The structure of the matrix $\mathbf{\Lambda}_{N \times N}$ is found to be the same for all values of $N$.
The quantities $\Gamma$ and $\tau$ are given by,
\begin{eqnarray}
\Gamma & = & \frac{w^2 + v^2}{E-\epsilon} \nonumber\\
\tau & = & \frac{v w}{E-\epsilon}
\end{eqnarray}

Two interesting points need to be paid attention to here. They are, 

$(i)$ The values of the quantities $\Gamma$ and $\tau$ appear to be the same for both the two-strand and the tree-strand case (as well as for the general $N$-strand case). \\
\noindent
$(ii)$ In the ladder network geometry, be it two strand or three-strand, the nearest, next-nearest (along the diagonals) and the next-next nearest neighbor hoppings have the same values $\Gamma$ and $\tau$. 
The second point above holds the key to working out the FB energy and the degeneracy, as will be clear now. 


\begin{figure}[ht]
\centering
\includegraphics[width=\columnwidth]{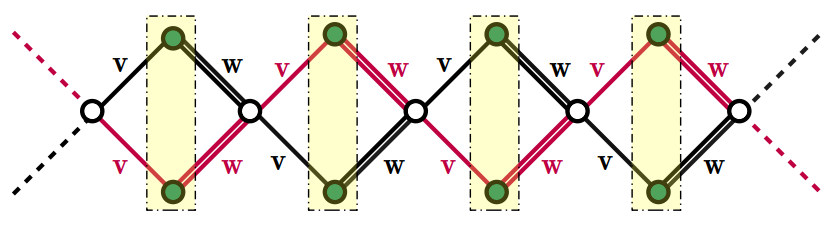}
\caption{(Color online) Amplitude distribution of the wave functions at an energy $E = 0$ for Fig.~\ref{fig1}(a).This configuration corresponds to the flat band and represents a compact localized state (CLS) ~\cite{leykam}. Here zero-amplitudes are marked by white color sites. The green colored sites indicate non-zero-amplitudes, normalized to $\pm 1$ so as to satisfy the difference equation, Eq.~\eqref{diffeqn}. The CLS locations are highlighted by yellow boxes.}
\label{cls1a}
\end{figure}


It is easily verified that, the potential matrix and the hopping matrix for the two-strand and the three-strand ladders commute, {\it independent of the energy} $E$. That is,
$[\tilde{\epsilon_{2}}, \tilde{t_{2}}]=0$ and $[\tilde{\epsilon_{3}},\tilde{t_{3}}]=0$, irrespective of the energy $E$. These matrices can therefore be diagonalized simultaneously using the same matrix $\mathcal{M}$ say, and the difference equations written in the matrix form above are easily written down in a new basis defined by $\Phi_n=\mathcal{M}^{-1}$ $\Psi_n$. The equations for each strand, in both cases are now completely decoupled in the new basis. For the two-strand ladder, the equations read, 
\begin{widetext}
\begin{eqnarray}
(E - \epsilon)  \phi_{n,1} & = & 0 \nonumber \\
\left (E - \epsilon - 2 \frac{v^2+w^2}{E-\epsilon} \right ) \phi_{n,2} & = &\frac{2 vw}{E-\epsilon}  (\phi_{n-1,2}+\phi_{n+1,2}) 
\label{twostrand}
\end{eqnarray}
\end{widetext}
Here $\phi_{n,j}$ the components of the column vector $\Phi_n$.

The first of the above pair of equations clearly show that at $E=\epsilon=0$ we should have an `atomic like' orbital. There is no `hopping' term on the right hand side, and so this state should be {\it localized} in character. This is actually a compact localized state (CLS)~\cite{leykam}, and is responsible for the flat, non-dispersive band in the $E-vs-k$ diagram. The presence of a single equation $E=\epsilon=0$ implies that this FB is non-degenerate. The amplitude profile for this CLS is displayed in Fig.~\ref{cls1a}. The green colored sites have $\psi_i =\pm 1$ consistent with the difference equation satisfied by the system.

The dispersive bands for this two-strand ladder network arise out of the second equation, and read, 
\begin{equation}
    E = \epsilon \pm \sqrt{2 (v^2+w^2) + 4 vw \cos~ka}
    \label{disp2strand}
\end{equation}
where, $k$ is the wave vector, and $a$ is the effective lattice constant of the $1$-d decoupled chains in the ${\Phi_n}$ basis.

Following a similar procedure, the three-strand ladder network can be decomposed into three separate chains, completely decoupled from each other when written in the $\Phi$ basis. The difference equations now read, 
\begin{widetext}
\begin{eqnarray}
(E -\epsilon) \phi_{n,1}  & = & 0 \nonumber\\
(E-\epsilon) \phi_{n,2} & = & 0 \nonumber\\
\left (E -\epsilon-3  \frac{v^2+w^2}{E-\epsilon} \right ) \phi_{n,3} & = & \frac{3 vw} {E-\epsilon} \left (\phi_{n-1,3}+\phi_{n+1,3} \right ) 
\label{threestrand}
\end{eqnarray}
\end{widetext}
From the first pair of Eqs.~\eqref{threestrand} we can easily discern a doubly degenerate FB at $E=\epsilon=0$. The two dispersive bands in this case are given by, 
\begin{equation}
  E = \epsilon \pm \sqrt{  3 (v^2+w^2) + 6 vw \cos ka}
    \label{disp3strand}
   \end{equation}

We have cross-checked the results by writing down the Hamiltonian in Eq.~\eqref{ham} in momentum space, viz, 

\begin{equation}
    H_i = \sum_k \psi_{k}^\dagger \mathcal{H}_{i}(\mathbf{k}) \psi_{k}
    \label{d3}
\end{equation}
where the subscript $i$ indicates different geometries in Fig.~\ref{fig1}. For example, in the case of two, three and four cross-linked SSH chains (Fig.~\ref{fig1}(a),(b),(c)) respectively, the kernels of the Hamiltonian are given by, 
\begin{widetext}
\begin{equation}
    \mathcal{H}_{1}(\mathbf{k}) = \begin{pmatrix}
\epsilon & v + w e^{-ika} & v + w e^{-ika}\\
v + w e^{ika} & \epsilon & 0\\
v + w e^{ika} & 0 & \epsilon\nonumber\\
\end{pmatrix}
\label{ham1}
\end{equation}

\begin{equation}
    \mathcal{H}_{2}(\mathbf{k}) = \begin{pmatrix}
\epsilon & v + w e^{-ika} & v + w e^{-ika} & v + w e^{-ika} \\
v + w e^{ika} & \epsilon & 0 & 0\\
v + w e^{ika} & 0 & \epsilon & 0\\
v + w e^{ika} & 0 & 0 & \epsilon\nonumber\\
\end{pmatrix}
\label{ham2}
\end{equation}

\begin{equation}
    \mathcal{H}_{3}(\mathbf{k}) = \begin{pmatrix}
\epsilon & v + w e^{-ika} & v + w e^{-ika} & v + w e^{-ika} &  v + w e^{-ika} \\
v + w e^{ika} & \epsilon & 0 & 0 & 0\\
v + w e^{ika} & 0 & \epsilon & 0 & 0\\
v + w e^{ika} & 0 & 0 & \epsilon & 0\\
v + w e^{ika} & 0 & 0 & 0 & \epsilon
\end{pmatrix}
\label{ham3}
\end{equation}
\end{widetext}

\begin{figure}[ht]
\centering
(a)\includegraphics[width=.8\columnwidth]{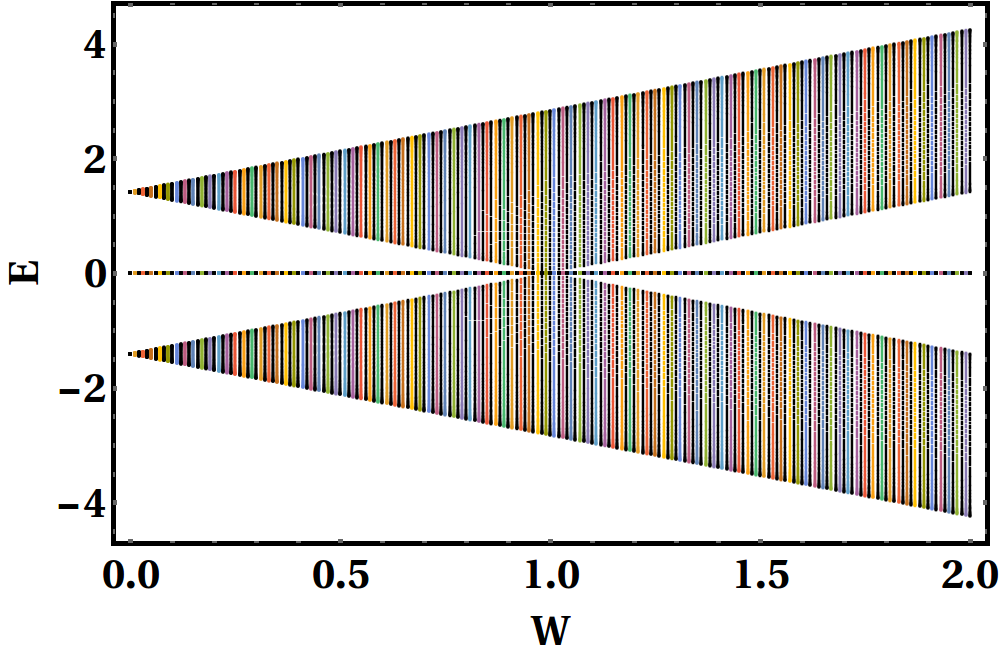}
(b)\includegraphics[width=.8\columnwidth]{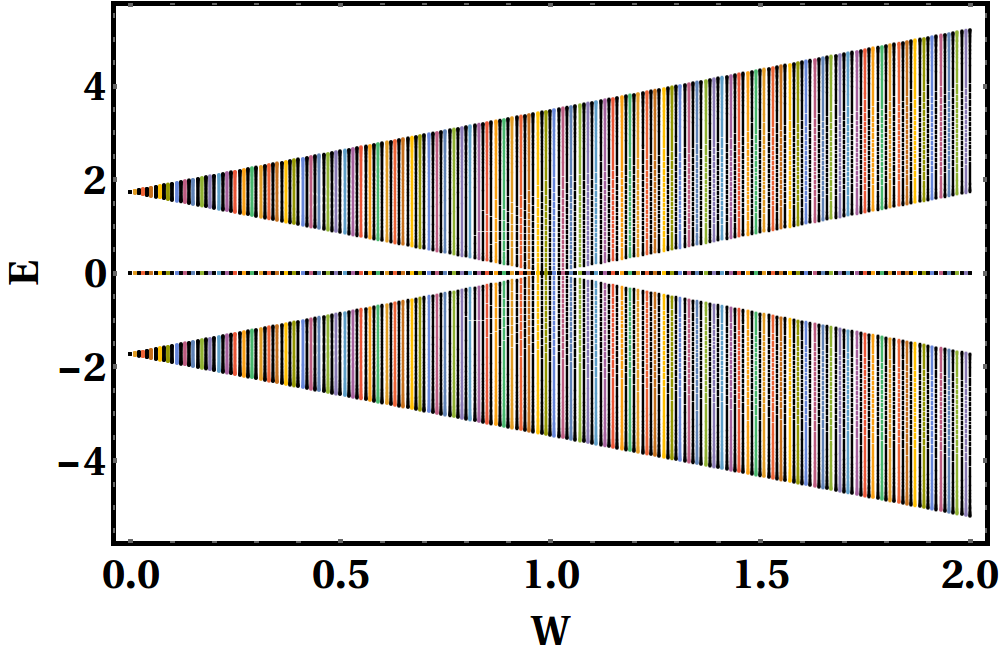}
(c)\includegraphics[width=.8\columnwidth]{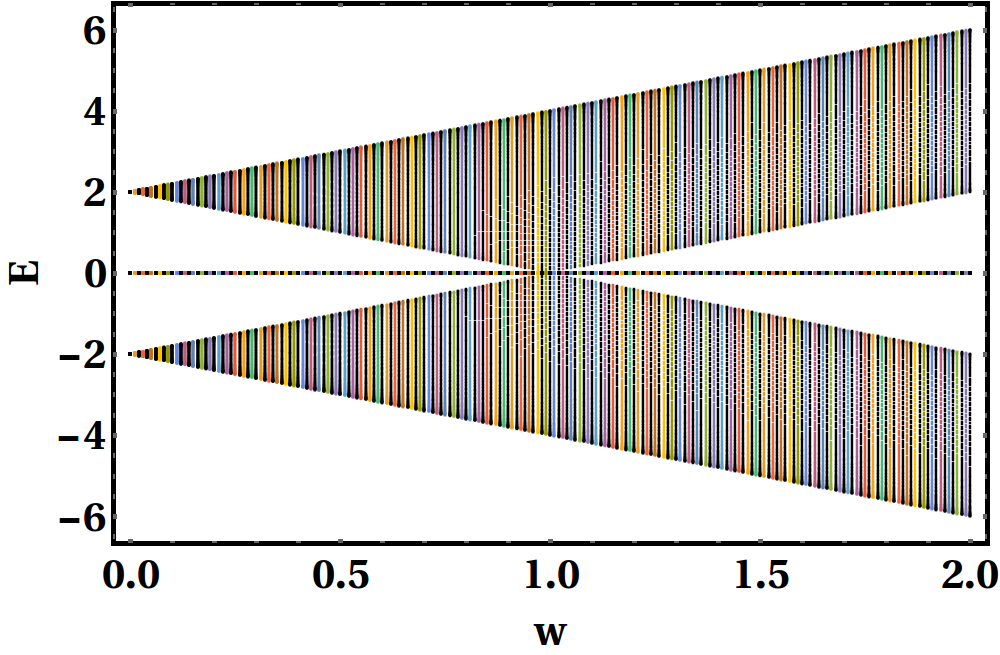}
\caption{(Color online) Distribution of energy eigenvalues of (a) two, (b) three and (c) four cross-linking SSH chains when the inter-cluster hopping $w$ varies. We have used open boundary conditions for $N_x=100$, where $N_x$ denotes the number of unit cells taken along the $x$-direction. The parameters are chosen as $ \epsilon = 0$ and $ v = 1$. The zero energy states are inherited from the flat bands, and are highly degenerate.}
\label{edgestates}
\end{figure}

\begin{figure*}[ht]
\centering
(a)\includegraphics[width=0.85\columnwidth]{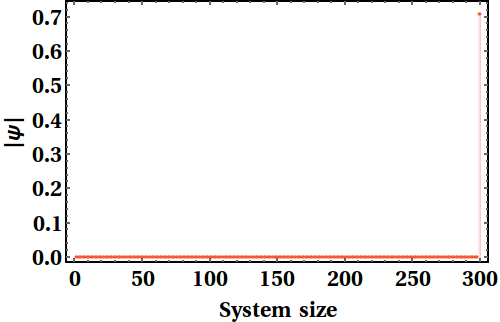}
(b)\includegraphics[width=0.85\columnwidth]{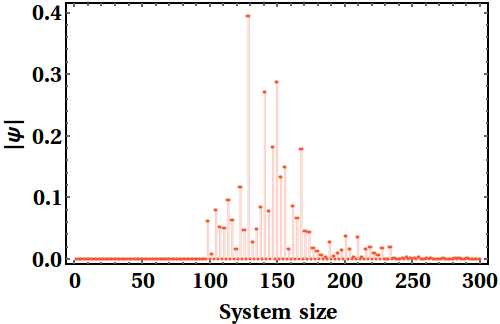}
(c)\includegraphics[width=0.85\columnwidth]{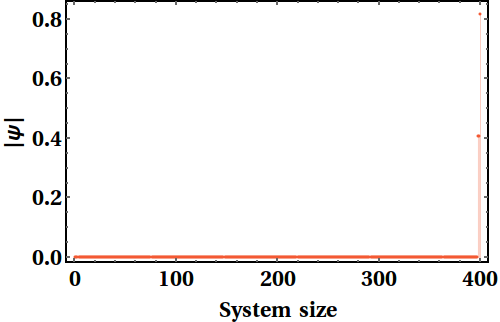}
(d)\includegraphics[width=0.85\columnwidth]{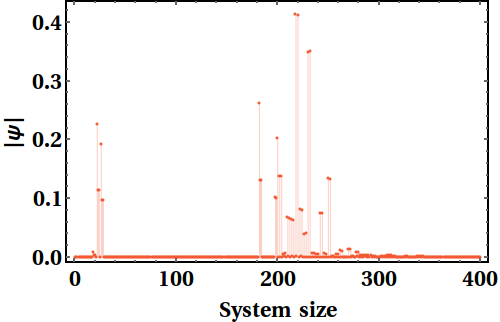}
(e)\includegraphics[width=0.85\columnwidth]{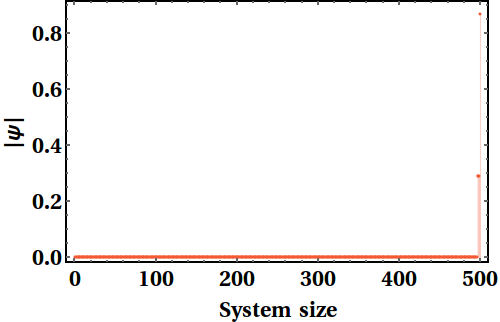}
(f)\includegraphics[width=0.85\columnwidth]{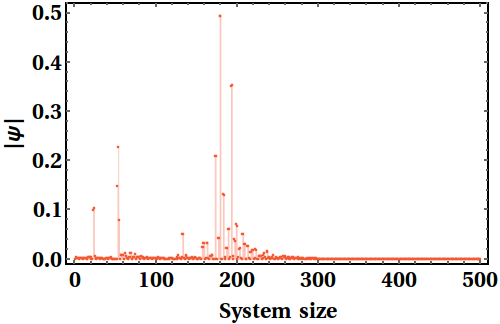}

\caption{(Color online)  Behaviour of zero energy wavefunction for  (a,b) two,  (c,d) three and (e,f) four entangled SSH arrays at {\it zero} energy, $E=0$. We have used open boundary conditions for $N_x=100$, where $N_x$ denotes the number of unit cells taken along the $x$-direction. The parameters are chosen as $ \epsilon = 0 $ and  $v = 1$, $ w = 1.1 $.}  
\label{ampli-edge}
\end{figure*}

The energy dispersion curves are displayed in Fig.~\ref{disp1} ($(a)$ - $(i)$) for different combinations of $v$ and $w$. The results of the first two figures are exactly reproduced by  Eq.~\eqref{disp2strand} and Eq.~\eqref{disp3strand}. 

Before we end this section, it must be mentioned that the formation of the compact localized states and the occurrence of the flat, non-dispersive bands is a result of the local topology of the underlying lattice structure, and is elegantly understood through the rank-nullity theorem~\cite{sutherland,lieb,matheus}. Our method is an alternative one.
\section{Topological properties}
\subsection{The Zak phase}

The closing and re-opening of an energy gap at the Brillouin zone boundaries are the primary signals for a topological phase transition (TPT). This needs to be substantiated by the change of the value of a topological invariant, such as the Zak phase ~\cite{zak}, from $0$ to a quantized value of $1$, for example.  The system is said to undergo a transition from a topologically trivial to a topologically non-trivial phase in such cases~\cite{asboth}. The Zak phase is purely a bulk property of the system, and therefore, we need to ensure the fulfillment of the  Born-von Karman periodic boundary condition. Some experiments in recent times, using photonic lattices~\cite{jiao} and cold atomic platform~\cite{bloch} have suggested mechanisms for a possible measurement of this topological invariant. We wish to examine whether the quantization of the Zak phase, as observed in a simple 1-d SSH model is still preserved under entanglement.

The Zak phase (in units of $\pi$) for the $n$-th bulk bands is defined as~\cite{asboth,bloch},

\begin{equation}
    Z = -i \oint_{BZ} \mathcal{A}_{nk} (k) dk
    \label{eq9}
\end{equation}
where, $\mathcal{A}_{nk}$ is the Berry curvature of the $n$-th Bloch eigenstate, and is given by~\cite{asboth},
\begin{equation}
    \mathcal{A}_{n{k}}(k)= \bra{\psi_{n{k}}}\ket{\frac{d\psi_{nk}}{dk}}
\end{equation}

The integral is performed along a closed loop in the Brillouin zone.  $\ket{\psi_{nk}}$ is the $n$-th Bloch state. We make use of the Wilson loop approach that is a gauge invariant formalism~\cite{fukui,wang}. It protects the numerical value of the Zak phase against any arbitrary phase change of Bloch wavefunction.

The integration in Eq.\ref{eq9} is converted into a summation over the entire Brillouin Zone, slicing it into $N = 240 $ identical discrete segments, each with a magnitude $\Delta k = 2 \pi/N$. The lattice constant $a$ is chosen to be unity. Convergence of the summation is assured with such a choice of $\Delta k$. The Zak phase for the non-degenerate $\alpha$-th band turns out to be,

\begin{equation} 
Z_\alpha = - Im ~ \left [ \log \prod_{k_n} \bra{\psi_{k_n,\alpha}} \ket{\psi_{k_{n+1},\alpha} } \right ]
\end{equation}
It is found that, for any value of $v < w$ the Zak phase for any dispersive band becomes exactly equal to unity, i.e. $Z=1$ (in units of $\pi$), implying a topologically non-trivial phase, while for $v > w$ we have $Z=0$ signifying a topologically trivial phase. This happens for any number of the cross-linking SSH chains so far we have checked. The exact similarity with a purely 1-d SSH chain, so far as the quantization of $Z$ is concerned, may be attributed to the fact that, in all the configurations depicting the entanglement, a travelling excitation will always find itself in a pure SSH environment even after crossing the nodal, cross-linking vertices. No matter how many SSH chains cross the nodes, the path of the hopping particle is still an unperturbed SSH chain. 

\subsection{The localized states at the edge and in the bulk}

The non-trivial points of difference in comparison with a one-dimensional SSH lattice arise when one considers a finite extent (along the $x$-direction) of the entangled SSH arrays and computes the distribution of the amplitudes $\psi_n$. The variation in the energy eigenvalues are shown in Fig.~\ref{edgestates} for two, three, and four cross-linking SSH chains. It is clearly seen that, right after the gap-closing condition $v=w$, a state in the energy gap appears for all $w > v$. This is analogous to the SSH scenario. However, there is a flat, non-dispersive band at $E=0$ that stems from the `looped' structure of the assembly of entangled SSH chains. The `gap-opening' energy $E=0$ coincides with the energy at which the FB appears for the bulk system. It is important to appreciate that the FB is truly a result of the preservation of the translational invariance, which is there as long as we focus on an infinitely large system. Truncating the lattice beyond a certain length gives us the scope to examine whether an eigenstate is localized at one of the edges. 

In Fig.~\ref{ampli-edge}  we have shown the edge states for two, three, and four cross-linking SSH chains. (a), (c) and (e) in Fig.~\ref{ampli-edge} clearly bring out the states localized at one edge of the systems ($100$ unit cells are taken here, in each case). The edge states are protected by chiral symmetry. The chiral symmetry operators for the three cases depicted in  Fig.\ref{fig1}(a),(b) and (c) have been found to be,  

\begin{equation}
\xi_{1} = \left[ \begin{array}{cccccccccccccccc}
-1 & 0 & 0\\
 0 & 1 & 0\\
 0 & 0 & 1\nonumber\\
\end{array}
\right ] 
\end{equation}

\begin{equation}
\xi_{2} = \left[ \begin{array}{cccccccccccccccc}
-1 & 0 & 0 & 0\\
 0 & 1 & 0 & 0\\
 0 & 0 & 1 & 0 \\
 0 & 0 & 0 & 1\nonumber\\
\end{array}
\right ] 
\end{equation}

\begin{equation}
\xi_{3} = \left[ \begin{array}{cccccccccccccccc}
-1 & 0 & 0 & 0 & 0\\
 0 & 1 & 0 & 0 & 0\\
 0 & 0 & 1 & 0 & 0 \\
 0 & 0 & 0 & 1 & 0\\
 0 & 0 & 0 & 0 & 1
\end{array}
\right ] 
\end{equation}
respectively. We have checked that the symmetry operator $\Gamma$ can be obtained for {it any} number of entangled SSH chains.
It can be easily checked that $\xi_{i}^{-1} \mathcal{H}_{i}(\mathbf{k}) \xi_{i} = -\mathcal{H}_{i}(\mathbf{k})$, for onsite potential $\epsilon = 0$. Where the subscript $i=1$, $2$, and $3$ correspond to the geometries in Fig.\ref{fig1} (a), (b) and (c) sequentially. We thus have symmetry-protected edge states for the geometries considered here, confirming the topological character of the entangled systems.

In the panels (b), (d) and (f) in Fig.~\ref{ampli-edge}, states localized at the bulk of the system appear for $E=0$. This is different from a simple one-dimensional SSH model, and the reason may be traced back to the existence of the {\it linking nodes}, where the coordination number changes as one changes the number of the chains that are cross-linked. This creates an environment of a {\it disorder}, and is responsible for these bulk-localized states. Each of the panels (b), (d) and (f) show three different localized states, picked up from the spectrum of the zero-energy bulk localized states corresponding to three different models. Needless to say, for any one specific model one could take a linear combination of such localized states, and still that combination would indeed be localized in the bulk of the system. But their properties will be sensitive to any kind of uncorrelated disorder introduced in the hoppings $v$ and $w$, while the edge-localized modes will remain unaffected, as is discussed below.

To check the robustness of the edge-localized modes and to see whether a change in the character of the bulk states affects the edge modes in any way, we introduced disorder in the distribution of the hopping amplitudes throughout the bulk as well as in the boundaries. The disorder was implemented by changing $v$ and $w$ in each unit cell to $v+\delta v$ and $w+\delta w$ respectively, where $\delta v$ and $\delta w$ were selected randomly from a window between $0$ and $1$. In all the cross-linked entanglement cases, we set $v=1$ and $w=1.1$.  Interestingly the appearance and the nature of the zero energy edge state remain unaffected, though the disorder will naturally change the localization properties of the bulk eigenstates, including the other `zero-energy' states in the bulk. The independence of the edge states (with the disorder) naturally rules out any interaction with the bulk localized states, including the zero-energy flat band. The edge states are therefore topologically protected. The bulk localized states do not relate to any topological feature of the system considered. They show up even when we choose $v > w$. The amplitude of a zero-energy state vanishes at the nodal points, and one such example is already shown in Fig.~\ref{cls1a}.
\section{Concluding Remarks}
We have analyzed the topological properties of cross-linked SSH chains, entangled through nodal points that are periodically placed in one dimension. The number of the SSH chains taking part in the geometry can be anything. The first interesting observation is the appearance of flat, non-dispersive bands that owe their origin to the looped structure. The degeneracy of these flat bands increases with an increasing number of SSH chains forming the cross-links. We give an analytically exact prescription to discern the energy eigenvalue corresponding to such a looped structure, and in addition to it, our prescription precisely unveils the degree of degeneracy. 

We have followed up the first part with a thorough investigation of the topological properties of an $N-$entangled system, and presented results here for $N=2$, $3$ and $4$. The systems exhibit topological properties, reflected through quantized Zak phase for all the Bloch bands, gap-states localized at one edge of the system at the gap-opening energy $E=0$. The bulk-boundary correspondence is honoured. The states have been found to be protected by chiral symmetry, and we have been able to work out the symmetry matrix in each case. In addition to this, and in contrast to a simple one-dimensional SSH chain, we also observe clusters of states localized in the bulk of the system - a fact that, to our minds, can be attributed to the flavour of a disordered environment, created at the cross-linking nodes.

\vskip .2in

\section{Acknowledgments}
S.B is thankful to the Government of West Bengal for the SVMCM (WBP221657867058) Scholarship.


\section{Statements and Declarations}

We confirm that the manuscript is the authors' original work and not copied or plagiarized version of some other published work.\\
\textbf{Data availability statement:}
Data will be available on request to the authors.\\
\textbf{Conflict of interest statement:}
We declare that this manuscript is free from any conflict of
interest. The authors have no financial interests or personal relationships that could have appeared to influence the work reported in this paper.\\
\textbf{Funding statement:}
No funding was received to support this work.\\
\textbf{Authors’ contributions:}
Sauvik Chatterjee developed theoretical concepts, solved mathematical parts analytically, prepared figures, analyzed the results, and wrote the manuscript. Sougata Biswas developed codes, prepared the figures and graphs, analyzed the results, and wrote the manuscript. Arunava Chakrabarti
conceptualized the problem, analyzed the results, and wrote the manuscript.\\

\end{document}